\RequirePackage{fix-cm} 
\documentclass[a4paper,twoside,reqno,11pt]{amsart}

\usepackage{fixltx2e}     
\usepackage{a4wide,verbatim}
\usepackage{hyperref}
\usepackage{amsmath,amsfonts,amssymb,amsgen,amsbsy,eucal,mathrsfs}

\newcommand{\ud}{\mathrm{d}}

\newcommand{\ue}{\mathrm{e}}

\newcommand{\vu}{\boldsymbol{u}}
\newcommand{\vx}{\boldsymbol{x}}

\newcommand{\half}{{\textstyle{1\over2}}}

\title{Remarks on Bernoulli constants, gauge conditions \\ and phase velocities in the context of water waves}
\author[D. Clamond]{Didier CLAMOND}
\newcommand{\nfont}{\fontshape{n}\selectfont}
\address{({\nfont\textbf{Didier Clamond}}) Universit\'e C\^ote d'Azur, CNRS-LJAD UMR 7351,
Parc Valrose, F-06108 Nice, France.} 
\email{didierc@unice.fr}

\begin{document}

\begin{abstract}
This short note is about the gauge condition for the velocity potential,
the definitions of the Bernoulli constant and of the velocity speeds in 
the context of water waves. These definitions are often implicit and thus 
the source of confusion in the literature. This note aims at addressing 
this issue. The discussion is related to water waves because the confusion 
are frequent in this field, but it is relevant for more general problems 
in fluid mechanics.  
\end{abstract}

\maketitle

\section{Introduction}\label{intro}

The Euler equations describe the momentum conservation of an inviscid fluid. 
For irrotational motions of incompressible fluids, the Euler equations can be 
integrated into a scalar equation called {\em Bernoulli equation} for steady 
flows and {\em Cauchy--Lagrange  equation} for unsteady flows. The Bernoulli 
and Cauchy--Lagrange  equations resulting of an integration procedure, they 
involve an arbitrary integration `constant', the so-called {\em Bernoulli constant} 
(that is actually an arbitrary function of time for the Cauchy--Lagrange  equation). 
This Bernoulli `constant' and its physical meaning is a frequent source of confusion 
in the literature, especially in the study of water waves. Thus, there are 
been some recent works aiming at clarifying the situation \cite{vasdec12}.

The purpose of this short note is to address the issues related to the Bernoulli 
constants. This leads to clarify the definition of the velocity potential, its uniqueness 
being introduced by a gauge condition, and how this quantity is modified via Galilean 
transformations. Various frames of references are also discussed as they lead to 
the definition the phase velocity of a wave.

\section{Cauchy--Lagrange  equation}

For the sake of simplicity, we consider the two-dimensional motion of an 
homogeneous incompressible fluid, but this is not a limitation for the purpose 
of the present note. For inviscid fluids, the equations of motion are \cite{sto}
\begin{align}
& u_x\ +\ u_y\ =\ 0, \label{eqinc}\\
& u_t\ +\ u\,u_x\ +\ v\,u_y\ =\ -p_x, \label{eqeulx} \\
& v_t\ +\ u\,v_x\ +\ v\,v_y\ =\ -p_y\ -\ g, \label{eqeuly} 
\end{align} 
where $\vx=(x,y)$ are the Cartesian coordinates ($y$ being directed upward), 
$t$ is the time, $\vu=(u,v)$ is the velocity field, $p$ is the pressure divided 
by the (constant) density and $g$ is the (constant) acceleration due to gravity 
directed toward the decreasing $y$-direction (downward).

In irrotational motion $v_x=u_y$, so there exists a velocity potential $\phi$ such 
that $u=\phi_x$ and $v=\phi_y$, i.e. $\vu=\mathrm{grad}\,\phi$. The Euler equations 
(\ref{eqeulx})-(\ref{eqeuly}) can then be rewritten \cite{sto}
\begin{equation}
\mathrm{grad}\!\left[\,\phi_t\ +\ \half\,(\phi_x)^2\ +\ \half\,(\phi_y)^2\ 
+\ g\,y\ +\ p\,\right]\,=\ 0,
\end{equation}
that can be integrated into the Cauchy--Lagrange  equation
\begin{equation} \label{eqlagcaugen}
\phi_t\  +\ \half\,(\phi_x)^2\ +\ \half\,(\phi_y)^2\  +\ g\,y\ +\ p\ =\ C(t),
\end{equation}
where $C$ is an integration `constant' often called {\em Bernoul\-li constant} 
or {\em Bernoul\-li integral}.

\section{Gauge condition}

The velocity potential being defined via its gradient, $\phi$ is not an unique 
function: adding any arbitrary function of time to $\phi$ does not change the 
velocity field. Thus, if one makes the change of potential \cite{sto}
\[
\phi(\vx,t)\ =\ \phi^\star(\vx,t)\ +\ \int C(t)\,\ud\/t,
\]
so that $\mathrm{grad}\,\phi=\mathrm{grad}\,\phi^\star$, the Cauchy--Lagrange  
equation (\ref{eqlagcaugen}) becomes
\begin{equation} \label{eqlagcaumod}
\phi^\star_t\ +\ \half\,(\phi_x^\star)^2\ +\ \half\,(\phi_y^\star)^2\ +\ g\,y\ +\ p\ =\ 0.
\end{equation}
In other words, this shows that it is always possible, via a suitable definition 
of the velocity potential, to take 
\begin{equation}\label{gauge}
C(t)\ =\ 0,
\end{equation}
without loss of generality and preserving the velocity field (i.e., 
$\mathrm{grad}\,\phi=\vu$). Enforcing the unicity of the velocity 
potential $\phi$ (up to an additional constant) via (\ref{gauge}) 
is a so-called {\em gauge condition}.

Hereafter, we always take the gauge condition (\ref{gauge}) and the 
Cauchy--Lagrange  equation is thus
\begin{equation} \label{eqlagcau}
\phi_t\ +\ \half\,(\phi_x)^2\ +\ \half\,(\phi_y)^2\ +\ g\,y\ +\ p\ =\ 0.
\end{equation}
Of course, other gauge conditions could be introduced, as well as no gauge 
condition at all. In the latter case, the arbitrary function $C(t)$ should 
be carried along all the derivations.

Note that with the gauge (\ref{gauge}), the Bernoulli `constant' disappears from 
the Cauchy--Lagrange  equation (\ref{eqlagcau}), but it has not completely been 
eliminated: it is now `hidden' in the definition of the velocity potential $\phi$ 
and will reappear explicitly for some special flows, as shown below.

\section{Galilean transformation}

Let be a change of Galilean frames of reference  $\mathcal{R}_0\mapsto\mathcal{R}_1$, 
where the coordinate system attached to $\mathcal{R}_1$ appears to travel at a constant 
speed $c$ in $\mathcal{R}_0$ along the $x$-direction. If $(x,y,t)$ and $(X,Y,T)$ denote 
the independent variables in $\mathcal{R}_0$ and $\mathcal{R}_1$, respectively, and if 
$(u,v,p)$ and $(U,V,P)$ are the corresponding velocity and pressure fields, the  Galilean 
transformation from $\mathcal{R}_0$ to $\mathcal{R}_1$ is \cite{LanLif1}
\begin{equation}\label{galtra}
X\, =\, x\, -\, c\,t, \quad  Y\, =\, y, \quad T\, =\, t, \quad U\, =\, u\, -\, c, \quad 
V\, =\, v, \quad P\,=\,p.
\end{equation}

From the Galilean transformation (\ref{galtra}), the transformation of the velocity 
potential $\phi\mapsto\Phi$ is necessarily (assuming $U=\Phi_X$ and $V=\Phi_Y$) of 
the general form
\begin{equation}\label{phigaltragen}
\phi\ =\ \Phi\ +\ c\,X\ +\ K(T),
\end{equation}
where $K$ is an arbitrary function of $T$ to be determined. Substituting (\ref{phigaltragen}) 
into the Cauchy--Lagrange  equation (\ref{eqlagcau}), one obtains at once
\begin{equation}\label{eqlagcaugalgen}
\Phi_T\ +\ \half\left(\Phi_X\right)^2\ +\ \half\left(\Phi_Y\right)^2\ +\ P\ 
+\ g\,Y\ =\ \frac{c^2}{2}\ -\ \frac{\ud\,K}{\ud\/T},
\end{equation}
which is also a Cauchy--Lagrange  equation. If we want to leave the Cauchy--Lagrange  
equation (\ref{eqlagcau}) invariant under Galilean transformations, the right-hand 
side of (\ref{eqlagcaugalgen}) must vanish, so one must take
\begin{equation}
K\ =\ \half\, c^2\,T,
\end{equation}
thence the Galilean transformation for the velocity potential: 
\begin{equation} \label{galtraphi}
\Phi\ =\ \phi\ -\ c\,x\ +\ \half\,c^2\,t, \qquad \phi\ =\ \Phi\ +\ c\,X\ +\ \half\,c^2\,T.
\end{equation}

The Galilean transformation (\ref{galtraphi}) for the velocity potential is such 
that it preserves the gauge condition (\ref{gauge}), i.e., the gauge condition 
(\ref{gauge}) is the same in all Galilean frames of reference if the velocity 
potential is transformed according to  (\ref{galtraphi}). 

Of course, it is not obligatory to choose the same gauge condition in every frame 
of reference. 
In that case, a suitable Galilean transformation should be introduced for the velocity 
potential, so that the gauge condition is transformed properly. However, it is simpler 
and less prone to confusion  to take the same gauge condition for all frames of reference. 
Note that the Galilean transformation of a velocity potential is similar to the 
one for the action in classical mechanics (see the problem at the end of \S8 in 
\cite{LanLif1}). 

The mishandling of the Galilean transformation for the velocity potential and the resulting 
(implicit) change of gauge condition is a source of apparent incompatibilities regarding the 
Bernoulli constants sometimes found in the literature. This confusion is also sometimes the 
source of physical misinterpretations as well. Discussion on this matter is given in the 
section \ref{secrem} below.

\section{Steady flow and Bernoulli equation} 

An important class of physical problems involve steady flows where all measurable quantities 
(velocity, pressure, density) are independent of the time $t$ in Eulerian description 
of motion.\footnote{The definition in Lagrangian description of motion can be found in 
\cite{cla2007}} A velocity potential is not directly measurable, only its gradient (i.e.,  
the velocity field) can be measured. The question is thus to find out the general form 
of a velocity potential for steady flows.

Since, for steady flows, $\mathrm{grad}\,\phi\/$ is independent of time, it follows 
that the most general potential of such flows is of the form
\begin{equation}
\phi(x,y,t)\ =\ \Phi(x,y)\ -\ A(t),
\end{equation}
where $A$ is a function of time only to be determined. Substituting this relation into 
(\ref{eqlagcau}), one gets
\begin{equation}\label{eqber0}
\half\,(\Phi_x)^2\ +\ \half\,(\Phi_y)^2\ +\ p\ 
+\ g\,y\ =\ \frac{\ud\,A}{\ud\/t}.
\end{equation}
The left-hand side of (\ref{eqber0}) being independent of the time $t$, so is 
the right-hand side. Therefore, $A$ is necessarily a linear function of $t$:
\begin{equation}
A(t)\ =\ \half\,B\,t\ +\ A_0,
\end{equation}
where $B$ is a Bernoulli constant (the factor $1/2$ is unessential and it is 
introduce only for later convenience). The Cauchy--Lagrange  equation thus 
becomes the Bernoulli equation
\begin{equation}\label{eqber}
\half\,(\Phi_x)^2\ +\ \half\,(\Phi_y)^2\ +\ p\ +\ g\,y\ =\ \half\,B.
\end{equation}
This shows that, for irrotational motions, the Bernoulli equation is just a 
special form of the Cauchy--Lagrange  equation without inconsistencies or 
contractions with respect of the Bernoulli constants. The key point is that 
the velocity potential is not independent of the time for steady flows under 
the gauge condition (\ref{gauge}).

\section{Traveling waves}

Consider now the more general case of a wave traveling at constant speed $c$ 
along the $x$-direction and without change of form. This concept can be made 
precise as follow.

A traveling wave is such that it exists a frame of reference where the flow 
appears steady. Thus, in any other Galilean frame of reference, the velocity 
potential of a traveling wave is, according to the sections above, necessarily 
of the form
\begin{equation} \label{teapot}
\phi(x,y,t)\ =\ \varphi(\xi,y)\ -\ \half\,\mathscr{B}\,t,
\qquad \xi\ =\ x\ -\ c\,t,
\end{equation}
where $\mathscr{B}$ is a Bernoulli constant. The Cauchy--Lagrange  equation 
thus yields the modified Bernoulli equation
\begin{equation}\label{eqbertra}
-c\,\varphi_\xi\ +\ 
\half\,(\varphi_\xi)^2\ +\ \half\,(\varphi_y)^2\ +\ p\ +\ g\,y\ =\ \half\,\mathscr{B}.
\end{equation}
If $c=0$ (frame of reference traveling with the wave) the equation (\ref{eqbertra}) 
yields the Bernoulli equation (\ref{eqber}). If $c\neq0$ then $\mathscr{B}\neq B$, 
as it can be easily seen considering the Galilean transformation from (\ref{eqber}) 
to (\ref{eqbertra}).

This result shows that, for a traveling wave, the velocity potential is 
generally not of the form $\phi(x,y,t)=\varphi(x-ct,y)$ but of the more general 
form (\ref{teapot}). The form of solution $\phi(x,y,t)=\varphi(x-ct,y)$ is, in 
most cases, incompatible with (\ref{gauge}) as it implicitly 
implies a different gauge condition for the velocity potential. As for steady flows, 
the violation of the gauge condition is the source of confusion  and apparent contradictions 
found in the literature.

\section{Phase velocities}

When looking for traveling waves propagating at constant speed $c$, one has 
to specify in which frame of reference the wave is observed. Otherwise 
confusion  and inconsistencies may occur. { Indeed, quite 
often in the literature, traveling wave solutions are sought as functions 
of $x-ct$, where $c$ is called the phase velocity in the `fixed' frame 
of reference without further precisions. This is not a definition of 
$c$ because there are infinitely many `fixed' Galilean frames. Below, 
we give the definitions of two `fixed' frame often used in practice.}

The fluid domain being $-d\leqslant y\leqslant\eta$ ($d$ the constant 
depth, $\eta$ the surface elevation from rest), there are two `fixed' 
frame of references commonly used. 
The frame of reference where the average horizontal velocity is zero 
at the bottom is defines such that
\begin{equation}\label{defrefce}
\frac{1}{T\,L}\int_{-T/2}^{T/2}\int_{-L/2}^{L/2} u(x,y\!=\!-d,t)\,
\ud\/x\,\ud\/t\ =\ 0,
\end{equation}
where $L$ and $T$ are, respectively, the wavelength and the period. 
The condition (\ref{defrefce}) defines univocally the phase speed. 
This is {\em Stokes' first definition of wave celerity} \cite{sto47}, 
sometimes denoted $c_\text{e}$ \cite{fen1990}. Since the flow is irrotational, 
this frame of reference is also the one where the horizontal velocity 
averages to zero along any horizontal line $y=\text{Constant}$ inside 
the fluid.

Another frame of reference of practical importance is the one where 
the mean flow is zero. The horizontal velocity is therefore such that
\begin{equation}\label{defrefcs}
\frac{1}{T\,L\,d}\int_{-T/2}^{T/2}\int_{-L/2}^{L/2}\int_{-d}^\eta 
u(x,y,t)\,\ud\/y\,\ud\/x\,\ud\/t\ =\ 0.
\end{equation}
The condition (\ref{defrefcs}) also defines univocally the phase speed. 
This is {\em Stokes' second definition of wave celerity} \cite{sto47}, 
sometimes denoted $c_\text{s}$ \cite{fen1990}.

In general $c_\text{e}\neq c_\text{s}$, but these two velocities 
are equal in deep water ($d\to\infty$) and for solitary waves ($L\to\infty$). 
If $B$ is the Bernoulli constant in the frame of reference moving 
with the wave (c.f. eq. \ref{eqber}), then $c_\text{e}^{\,2}\neq B\neq 
c_\text{s}^{\,2}$ in general but $B=c_\text{e}^{\,2}=c_\text{s}^{\,2}$ 
in deep water and for solitary waves \cite{cladut17}.

Many other frames of reference can of course be defined, their interest 
depending on the problem at hand. Moreover, one can use different average 
operators than the ones considered above (arithmetic mean in Eulerian description 
of motion). These remarks are also valid, of course, for unsteady flows. 
In any case, precise definitions are necessary to avoid confusion.

For practical determination of travelling waves, it is simpler to first 
determine the solutions for $\Phi$ and $B$ in the frame of reference moving with the wave 
where the flow is steady. In particular, since $\eta$ is the surface elevation from rest 
where the pressure is zero, the Bernoulli constant can be obtained averaging the Bernoulli 
equation at the free surface, i.e. 
\begin{equation*}
B\ =\ \frac{1}{L}\int_{-L/2}^{L/2} \left[\,\mathrm{grad}\,\Phi\,\right]^2_{y=\eta}\,\ud\/x.
\end{equation*}
At this stage, no phase speed needs to be defined. The phase speed 
$c$ is subsequently obtained for whatever frame of reference of interest, such as 
(\ref{defrefce}) or (\ref{defrefcs}). Thus, in the frame moving with the wave, the 
celerities $c_\text{e}$ and $c_\text{s}$ are computed with the formulae 
\begin{equation*}\label{defrefcecs}
\frac{1}{L}\int_{-L/2}^{L/2} \Phi_x(x,y\!=\!-d)\,\ud\/x\ =\ -c_\text{e}, \qquad
\frac{1}{L\,d}\int_{-L/2}^{L/2}\int_{-d}^\eta \Phi_x(x,y)\,\ud\/y\,\ud\/x\ =\ -c_\text{s}.
\end{equation*}
Finally, the velocity potential $\varphi$ and the Bernoulli constant $\mathscr{B}$ are 
obtained in the `fixed' frame of reference using the Galilean transformation 
(\ref{galtraphi}).

It should be noted that for experiments in closed flumes, the fixed (laboratory) 
frame of reference is the one without mean flow. But since viscous fluids (generally 
water) are used in practice, the fluid sticks to rigid walls and the velocity 
is zero at the bottom. Thus, the fixed frame is also the one without mean velocity 
at the bed, in contraction with the potential flow theory. This example shows that     
precise comparisons with experiments are not an easy matter.

\section{Discussion}\label{secrem}

Apparent incompatible formulations of the Bernoulli equations found in the 
literature (e.g. in \cite{ConStr2010,DecOli2011}) has been investigated by Vasan and Deconinck 
\cite{vasdec12}. They have shown that under the gauge condition (\ref{gauge}), considering 
$\phi(x-ct,y)$ for traveling waves yields (\ref{eqbertra}) with $\mathscr{B}=0$, that is generally 
incorrect. They solve this problem reintroducing a Bernoulli constant and they show that the 
resulting equation corresponds to another velocity potential. This velocity potential 
corresponds to a different gauge condition. Vasan and Deconinck \cite{vasdec12} also discuss the 
relation between Bernoulli constants and uniform currents. Such considerations are not 
necessary if traveling waves are defined by (\ref{teapot}) and if Galilean transformations 
between velocity potentials are defined by (\ref{galtraphi}) in order to preserve the 
gauge condition.  

In many applications, these apparent incompatibilities are of little consequence since they 
can be circumvented via redefinition of some parameters and variables, as shown in \cite{vasdec12}. 
In some applications, however, the consequences can be dramatic and no simple redefinitions 
can be introduced to solve the issue. 

An example is the case of a varying dissipation used 
to model sponge layers described in section \S4.2.2 of \cite{cla2005}. Indeed, in order to 
introduce a sponge layer, the Cauchy--Lagrange equation (\ref{eqlagcaugen}) is modified as
\begin{equation} \label{eqlagcaugendis}
\phi_t\  +\ \half\,(\phi_x)^2\ +\ \half\,(\phi_y)^2\  +\ g\,y\ +\ p\ +\ \gamma(x,y)\,\phi
\ =\ C(t),
\end{equation}
where $\gamma>0$ in the regions where damping is required and $\gamma=0$ elsewhere. If 
$\gamma$ is constant, the Bernoulli integral $C(t)$ can be set to zero without loss of 
generality via the change of potential
\[
\phi(\vx,t)\ =\ \phi^\star(\vx,t)\ +\ \int_{t_0}^t \ue^{\gamma(t'-t)}\,C(t')\,\ud\/t'.
\]
When $\gamma=\gamma(x,y)$, the gauge condition $C(t)=0$ cannot be applied because 
\[
\mathrm{grad}\,\phi\ =\ \mathrm{grad}\,\phi^\star\ +\ (\mathrm{grad}\,\gamma)
\int_{t_0}^t (t'-t)\,\ue^{\gamma(t'-t)}\,C(t')\,\ud\/t',
\] 
so equation (\ref{eqlagcaugendis}) is modified. One can easily verify that no transformation 
$\phi\to\phi^\star$ with $\mathrm{grad}\,\phi\,=\,\mathrm{grad}\,\phi^\star$ leads to 
(\ref{eqlagcaugendis}) with $C=0$ if $\gamma$ is not constant. Therefore, if one uses 
(\ref{eqlagcaugendis}) together with $C=0$ then $\mathrm{grad}\,\phi$ is not the 
velocity field and spurious unphysical phenomena may appear as discussed in \cite{cla2005}.

\section{Conclusion}

We have seen that the confusion regarding the Bernoulli constants can be the 
result of incorrect Galilean transformations violating the Gauge condition 
used for the velocity potential. This can also leads to incorrect mathematical 
definitions of the velocity potential for steady flows and traveling waves. 

Although the focus was on irrotational water waves, such considerations also 
apply for rotational waves \cite{Dyn2000} and for ideal fluid motions in general 
\cite{kam2003}. Remarks on the possible gauge conditions for variants of the 
Cauchy--Lagrange  equation involving varying dissipative terms are also discussed. \\

{\bf Acknowledgement.} The author is grateful to Dr Vishal Vasan for helpful discussions.

This material is based upon work supported by the National 
Science Foundation under Grant No. DMS-1439786 while the author was in residence 
at the Institute for Computational and Experimental Research in Mathematics in 
Providence, RI, during the Spring 2017 semester.

\end{document}